\documentclass[floatfix,twocolumn,amsmath,amssymb]{revtex4}
\PassOptionsToPackage{fleqn}{amsmath}

\usepackage{graphicx}
\usepackage{dcolumn}
\usepackage{bm}
\usepackage{color}

\begin{document}

\title{The ``dark phase'' in Sr$_2$Ir$_{1-x}$Rh$_x$O$_4$ revealed by Seebeck and Hall measurements}

\author{L. Fruchter and V. Brouet}%
\affiliation{Laboratoire de Physique des Solides, C.N.R.S., Universit\'{e} Paris-Sud
\\ Universit\'{e} Paris-Saclay, 91405 Orsay Cedex, France}

\begin{abstract}{}
It was found that, although isovalent, Rh substituted for Ir in Sr$_2$IrO$_4$ may trap one electron inducing effective hole doping of Ir sites. Transport and thermoelectric measurements on Sr$_2$Ir$_{1-x}$Rh$_x$O$_4$ single crystals presented here reveal the existence of an electron-like contribution to transport, in addition to the hole-doped one. As no electron band shows up in ARPES measurements, this points to the possibility that this hidden electron may delocalize in disordered clusters.
\end{abstract}
\maketitle

In the perovskite coumpound Sr$_2$IrO$_4$ - structurally similar to the first discovered cuprate superconductor, (La,Ba)$_2$CuO$_4$ - the strong spin-orbit coupling (SOC) allows for an effective localized state, entangling spin and orbital degrees of freedom, with total angular momentum $J_{eff} = 1/2$. The extended 5d orbitals reduce the electron-electron interaction, but this is compensated by the lifting of degeneracy induced by SOC. The resulting spin-orbital insulating state was early proposed to be the analog of the Mott insulating state found in cuprates\cite{Kim08,Jackeli09}. Doping the insulator either by oxygen vacancies\cite{Korneta10,Fruchter18} or rare earth substitution\cite{Klein08,Ge11,Brouet15} can be done only within a limited range, as compared to cuprates, and possible superconductivity was reported for surface doping only up to now\cite{Yan15}. The isotructural Sr$_2$RhO$_4$ being a paramagnetic metal\cite{Baumberger06,Martins11}, this motivated the systematic investigation of Sr$_2$Ir$_{1-x}$Rh$_x$O$_4$ in the entire doping range, revealing an antiferromagnetic insulator to a paramagnet transition at x $\approx$ 0.16, associated to a strong decrease of the resistivity\cite{Qi12,Clancy14}. While Sr$_2$RhO$_4$ is a metal due to the reduced SOC as compared to Sr$_2$IrO$_4$, the early proposal of an insulator to metal transition (IMT) essentially driven by SOC reduction, while Rh and Ir would have the same valence\cite{Klein09,Qi12,Lee12}, revealed inadequate, and Ir$^{4+}$ is actually substituted by Rh$^{3+}$ at small x value \cite{Klein08,Clancy14,Sohn14,Cao16}, as was confirmed theoretically \cite{Liu16}.

Transport studies in Ref.~\onlinecite{Qi12} showed a six order magnitude drop and a metallic temperature dependence of the resistivity for $x = 0.07$. In Ref.~\onlinecite{Cao16}, it was found that the chemical potential shifts to the lower Hubbard band at $x <0.05$ , while doping with $x= 0.11$  yields a metallic resistivity. Optical spectroscopy\cite{Xu20} indicates a collapse of the Mott gap at $x = 0.055$ and the appearance of a Slater gap (opening below $T_N$) with a Drude peak at $x = 0.07$. As doping directly occurs into the Ir-O$_2$ planes of this two-dimensional material, whether the charges induced by Rh are localized or not is essential for doping. In a local description, it was argued that the charge is first localized on nearby Ir$^{5+}$ site, the transition to the paramagnet being a consequence of 2D bond percolation\cite{Clancy14,Liu16}. When the nearby Ir$^{5+}$ site coordination number is less than a critical value, Rh valence was assumed to be +4\cite{Chikara17}, so that the average Rh valence continuously increases from +3 to +4 with doping, with Rh$^{+3}$ $\approx$ 80\% at $x = 0.15$. Besides this local description for doping, ARPES observed a homogeneous and rigid shift of the band structure towards the Fermi level \cite{Brouet15,Cao16,Louat18}. The Fermi surface defines a hole pocket containing $x$ holes up to x=0.15 \cite{Louat18}, implying that the holes delocalize on the Ir-O$_2$ lattice. A pseudogap is found at the Fermi level \cite{Cao16,Louat18},  which was tentatively attributed to disorder\cite{Louat19}. This is a situation different from the hole doped Sr$_{2-x}$K$_x$IrO$_4$, for which ARPES on thin films revealed a complete collapse of the Mott gap and a Fermi surface without pseudogap. While holes would be expected for Rh doping from these previous experiments, our study reveals that electron-like carriers nevertheless show up in the transport properties.\newline

\begin{figure}
\resizebox{0.9\columnwidth}{!}{%
\includegraphics{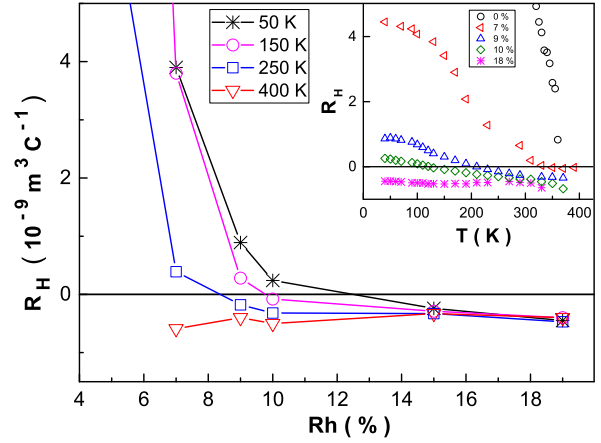}
}
\caption{Hall number.}\label{RHT}
\end{figure}

Single crystals of Sr$_2$Ir$_{1-x}$Rh$_x$O$_4$ were grown with a flux method\cite{Kim09,Louat18}. Their Rh-content was determined by energy dispersion x-ray (EDX) analysis, and found close to the nominal composition. No phase separation was observed from the EDX analysis. Magnetization data showing the decrease of $T_N$ with doping was provided in Ref.~\onlinecite{Xu20}. The crystals were typically 1 mm long and a few hundred $\mu$m thick. Longitudinal transport and thermoelectric power measurements were performed along the natural growth direction (110), \textit{i.e.} along the Ir-O-Ir bound. The thermoelectric measurements were performed using two Constantan-Chromel couples about 0.5 mm apart, with an AC thermal gradient, typically 100 K~m$^{-1}$. The absolute Seebeck coefficient for the couple metals were calibrated using a platinum reference.

\begin{figure}
\resizebox{0.9\columnwidth}{!}{%
\includegraphics{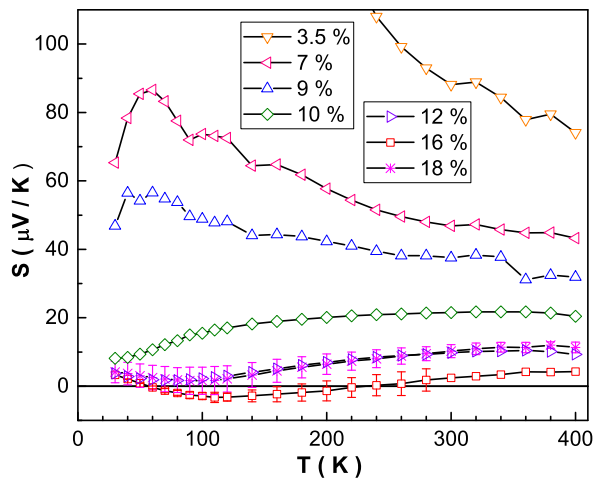}
}
\caption{Seebeck coefficient. The error bars for $x = 0.16$ and $x = 0.18$ are estimated from the data spread for two distinct samples.}\label{S}
\end{figure}

Hall measurements reveal a striking, unexpected feature: while it is positive for the low Rh dopings, as always observed in pure Sr$_2$IrO$_4$ (ref), it becomes negative at high temperature when the doping increases, and even remains entirely negative for $x \gtrsim$ 0.10-0.12 (Fig.~\ref{RHT}). All data seem to converge to a doping independent value at high temperature. The Seebeck coefficient ($S$) shows two regimes: for $x \lesssim$ 0.1, it displays large, positive values, decreasing with temperature, whereas $S(T)$ is less than 20 $\mu$V~K$^{-1}$ for larger doping values (Fig.~\ref{S}). The low doping behavior is typical of the one for an insulator, as observed in semiconductors and undoped cuprates\cite{Herring54,Ishii87}. At higher doping, the smaller value of $S(T)$ and the linear slope at low temperature are indicative of a metal\cite{Dugdale95}. The slope is positive at $x = 0.1$ and negative for $x \geq 0.12$, suggesting hole and electron carriers, respectively. Within error bars, $S(T)$ is found negative at $x = 0.16$, which appears to arise from the negative slope at low temperature for $x \geq 0.12$. However, the evolution is non-monoteneous with increasing doping, as $S(T)$ for $x = 0.18$ is again positive, in a re-entrant fashion.

Therefore, both transport and thermoelectric data indicate that there is an electron-like contribution to transport at the higher doping levels, whereas it is usually considered as a hole-doped system. For the highest doping ($x = 0.18$), we observe that the Hall number is temperature-independent with $R_H$ $\simeq$ - 4.5 10$^{-10}$ m$^3$~C$^{-1}$, \textit{i.e.} 1.4 e$^{-}$ per Ir/Rh site. As this is close to unity, a first possibility would be that hole-doping has induced a collapse of the Mott gap, so that all carriers from the Ir site contribute to transport; not only the $x$ doped ones. Such a transition was observed by Hall measurements in cuprates \cite{Badoux16}, and recently by ARPES in Sr$_2$IrO$_4$ films hole-doped with K \cite{Nelson20}. The corresponding Fermi Surface would be a simple circle for $x = 0.2$. Yet, ARPES clearly identify hole pockets at this doping with a residual gap of 0.1 eV along $\Gamma$M at $x = 0.16$ \cite{Louat18}. Moreover, the global evolution of the Hall constant points to a coexistence of holes and electrons, the former dominating at low temperatures up to $x = 0.1$. The number of holes at low temperatures would be unrealistically large in this limit, if it was considered originating from a single carrier (2.8 hole/Ir, for $x$ = 0.1). This calls for a two-fluid description for transport, which is rather natural as Rh creates both one hole and one electron. The hole, residing mostly on Ir site, is not perfectly delocalized as evidenced by the ARPES pseudogap, and the electron, believed to be essentially localized at the Rh sites, could hop from site to site when the number of Rh increases. 

To gain more insight into this coexistence, we use the conventional expression for the Hall effect for two-carrier Drude model\cite{Ashcroft}: 

\begin{equation}
R_H = \frac{1}{\alpha e} \frac{x_o + x_h-\beta^2 \, x_e}{[x_o + x_h+\beta \, x_e]^2}
\label{RHsim}
\end{equation}

where $x_o$ is the initial hole concentration \textit{per} Ir atom (of the order of 0.1 \%, Ref.~\onlinecite{Klein09}), $x_h$ and $x_e$ are the hole and electron doping fraction, $\beta$ = $\mu_e/\mu_h$ is the mobility ratio of carriers, and $\alpha$ = 9.8 10$^{27}$ m$^{-3}$ is the Ir concentration in Sr$_2$IrO$_4$.

\begin{figure}
\resizebox{0.9\columnwidth}{!}{%
\includegraphics{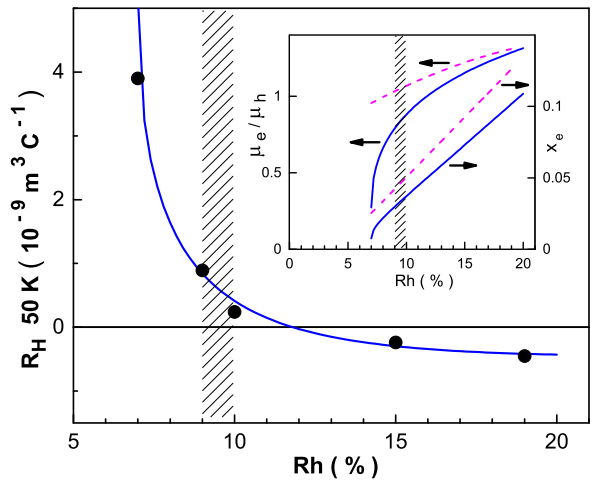}
}
\caption{Hall constant at 50 K, the full line indicating the simulation obtained applying eq. 1 with parameters shown in the inset. The hatched area indicates the crossover from insulating to correlated regime from the thermoelectric measurements. The inset displays the mobility ratio and the electron doping induced fraction needed to account for the data, using equation \ref{RHsim}, with $x_o = 6 \, 10^{-4}$, $x_e = x$ and $\beta = 1.5$ respectively. The full curves are for $T$ = 50 K and the dashed ones for $T$ = 250 K.}\label{simul_RH}
\end{figure}

We assume a number of holes $x_h$ $\simeq$ $x$, as measured  for $x$ $\lesssim$ 0.2 \cite{Chikara15,Louat18}. Then, we consider two different situations. In the first one, one hole and one electron \textit{per} Rh atom are induced by doping ($x_h = x_e = x$), but they may have different mobilities ($\beta$ is doping dependent). In the second one where $\beta$ is a constant, but $x_e$ is doping dependent. To account for our data at the lowest temperature, this yields either a mobility ratio which drops at $x_c$ $\simeq$ $0.07$, or an electron concentration rising linearly, as $x_e$ $\simeq$ $\gamma \,(x-x_c)$, where $\gamma = 1$ for $\beta = 1.5$ and lower for larger $\beta$ values (Fig.~\ref{simul_RH}). Both results actually essentially account for the same situation, where electrons are either localized or absent below $x_c$. It is clear from the data in Fig.~\ref{RHT} that thermal activation promotes electron doping, which shows up either in the computed mobility ratio or in electron concentration at high temperatures, depending on the model (Fig.~\ref{simul_RH}, inset). This is in favor of the existence of frozen electron carriers below $x_c$.

We favor a two-phase scenario, where part of the charge of the Ir$^{+5}$/Rh$^{3+}$ defects may be delocalized on the Ir-O$_2$ lattice as holes, and electrons dope the Rh-O$_2$ lattice of Rh clusters. In this scenario, the Seebeck coefficients of the two phases would be mixed roughly in proportion of the phase volume, as is known for 2D binary composite materials\cite{Levy92,Kleber05}. Then, the evolution with doping of the Seebeck coefficient magnitude for each phase could yield a non-monotonous variation for the composite. This two-phase scenario would be consistent with the observation of a re-entrant behavior for $S(T)$, which is on the contrary difficult to reconcile with the growth of an electron pocket with doping.

\begin{figure}
\resizebox{0.95\columnwidth}{!}{%
\includegraphics{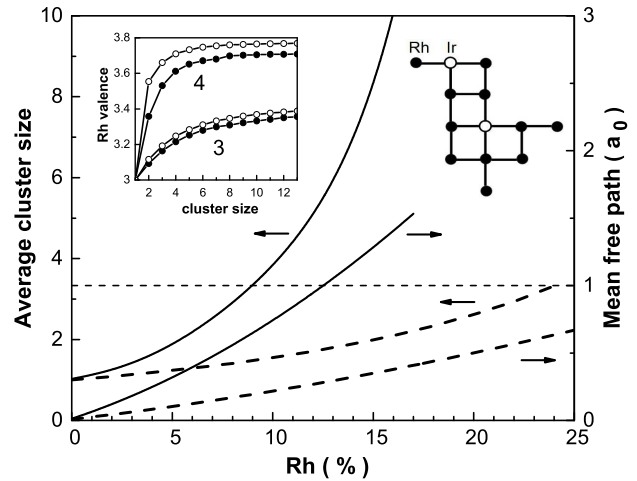}
}
\caption{Full lines: Size and mean free path for Rh clusters including Ir sites (as depicted). Dashed lines: pure Rh clusters. Dotted line at $l = 1$ is a Mott-Ioffe-Regel criterion for metallicity. Inset: average Rh valence, assuming critical Ir coordination 3 or 4, as in Ref.~\onlinecite{Chikara17} (full symbols: $x = 0.12$, open ones: $x = 0.18$). }\label{cluster}
\end{figure}

We now examine the expected structure of such clusters, and their possible contribution to the thermopower. To contribute to transport, the charges on the clusters must be delocalized. We expect this to happen when the mean free path exceeds a critical length, as given by a Mott-Ioffe-Regel criterion: $l \approx$ $a_o$, where $a_o$ is the Ir-O$_2$ lattice constant (we assume that thermal activation or external bias renders the Kubo gap irrelevant here\cite{Johnston02}). We further assume that the mean free path is given by the size of the cluster. Applying this criterion to conventional Rh clusters, where two Rh atoms belong to the same cluster when they share an Rh-O-Rh bond, would yield an average metallic cluster only for $x \gtrsim$ 0.32, \textit{i.e.} about half the critical percolation concentration ($x = 0.59$\cite{Stauffer91}) (Fig.~\ref{cluster}), which is well out of the investigated range. To account for metallic clusters at lower doping, we need to extend the definition of the clusters, and allow for Rh-Ir-Rh conductive bridges, as depicted in Fig.~\ref{cluster}. This brings the metallic transition of clusters down to $x \simeq$ 0.12, and the percolation threshold down to $x \simeq$ 0.3. In Ref.~\onlinecite{Clancy14}, the disappearance of antiferromagnetism with doping was interpreted as a percolation mechanism, at $x \simeq$ 0.2. Although we consider here an additional criterion for charge delocalization and make no hypothesis on the existence of long range antiferromagnetism outside of the clusters, we note that Ref.~\onlinecite{Clancy14} interpretation also relies on the incorporation of Ir$^{5+}$ atoms to a percolating, non-magnetic Rh cluster, which lowers by a factor 2 the critical concentration.

The doping levels of these clusters is very much dependent upon their size. Assuming Rh valence rule as in Ref.~\onlinecite{Chikara17} (Rh$^{3+}$ for isolated Rh and Rh$^{4+}$ otherwise), it is clear that doping is essentially provided by insulated Rh atoms, and that larger clusters ($l \geq a_o$) exhibit a larger Rh valence. This is the reason why doping becomes exhausted when $x \gtrsim 0.2$\cite{Chikara17,Louat18}. As a result, the large clusters are far from being doped with one charge per Rh atom, and so are not expected to be insulating. We have plotted in Fig.~\ref{cluster} (inset) the average valence for Rh upon cluster size, for critical Ir coordination 3 and 4.

\begin{figure}
\resizebox{0.9\columnwidth}{!}{%
\includegraphics{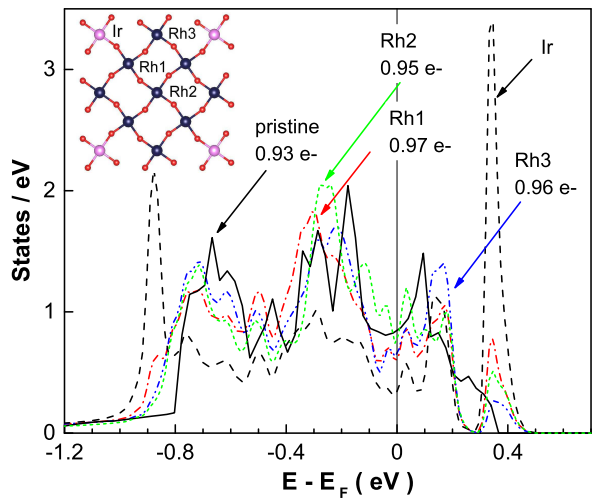}
}
\caption{Total projected orbital atomic Rh and Ir density of states for the supercell plane with Ir, as depicted. The electron number is the integrated density from -1.2 eV to the Fermi level. The pristine Sr$_2$RhO$_4$ Rh DOS is shown as full line.}\label{dos}
\end{figure}

To substantiate the existence of metallic Rh clusters incorporating Ir atoms, we performed DFT computations on a Sr$_2$Rh$_{(1-x)}$Ir$_x$O$_4$ supercell. For the pure Rh compound, it was shown that DFT, including SOC and on-site repulsion (DFT+SOC+U), correctly accounts for the experimental band structure\cite{Liu08,Martins11}, and we extend these to the case where $x$ is low. Considering the quasi-two-dimensional nature of the pure compound\cite{Baumberger06}, the supercell Sr$_2$Rh$_{0.935}$Ir$_{0.0625}$O$_4$, where, in one Rh-O$_2$ plane over two, one Rh atom over 8 is substituted with Ir (Ir atoms are separated by 3 Rh atoms, Figure ~\ref{dos}) is an appropriate model for an insulated Ir impurity in an infinite quasi-two-dimensional Rh cluster. We computed the electronic states using the full-potential linearized augmented plane wave (FP-LAPW) method, as implemented in the Wien2k software\cite{Blaha90}. As $U$ for Ir spans between 1.6 eV \cite{Liu16} and 2 eV \cite{Martins11,Chikara15}, and for Rh a smaller value between 1.2 and 1.4 eV \cite{Liu16,Ahn15,Liu08} (at odds with considerations for a smaller 4d Rh\cite{Sasioglu11}), we used $U_{\texttt{Ir}}$ = 1.6 eV and $U_{\texttt{Rh}}$ = 1.2 eV\cite{Liu16}. As $U_{\texttt{Ir}} > U_{\texttt{Rh}}$, the on-site repulsion is not expected to stabilize an insulating state.

As may be seen from the Rh density of states (DOS) (Figure~\ref{dos}), the number of states of the conduction band below $E_F$ differs from the one in the pristine material by $\delta$ = +0.02-0.04 $e^-$ per Rh, \textit{i.e.} +0.2 $e^-$ per Ir, summing up all Rh states. This indicates that there is a small electron doping effect from the Ir impurity, which is in line with Refs.~\onlinecite{Liu16,Chikara15}, where average valence state is far from Ir$^{+5}$. Unlike what was calculated in Rh doped Sr$_2$IrO$_4$\cite{Liu16}, the doping is here not restricted to first neighbors. The DOS for Ir shows that the Ir states overlap strongly with the Rh ones at the Fermi level, indicating that the Ir states contribute to the conduction band. This is also a different situation from the one when Sr$_2$IrO$_4$ is doped with Rh, where a narrow impurity band first appears inside the gap\cite{Chikara15}.

In this dilute Ir example, the charge transfer per Rh atom is small, as expected when the number of nearest Ir neighbors is small, leaving Rh essentially in the valence state +4. We expect that such large clusters provide a metallic, positive contribution to $S(T)$. This would be consistent with the re-entrant behavior for $x = 0.18$. It was indeed found that the thermopower for electron-doped Sr$_{2-x}$La$_x$RhO$_4$ does not substantially deviate from the one of a metal with holes until $x \lesssim$ 0.5, above which an insulator is formed\cite{Shimura94,Kwon19}. As underlined above, the Rh valence is expected to be lower in smaller clusters (Fig.~\ref{cluster}), opening the possibility for electron-doped ones. Tentatively assuming that the Seebeck coefficient for each phase contributes roughly in proportion of the phase volume\cite{Levy92,Kleber05}, we can evaluate these coefficients from a couple of $S(T)$ data. Doing so for $x = 0.10$ and $x = 0.16$ (with phase volume 0.12 and 0.21, for clusters as in Fig.~\ref{cluster}), we obtain an electron-doped contribution as large as -150 $\mu$V~K$^{-1}$. This is in apparent contradiction with the fact that Sr$_{2-x}$La$_x$RhO$_4$ phase with $x = 0.8$ is still an insulator with holes\cite{Shimura94,Kwon19}. Such macroscopic arguments must however be considered here with caution, as one deals with nanoscale structures close to the localized regime (Fig.~\ref{cluster}), and doping from in-plane Ir is likely very different from the one of the out-of-plane La.

To summarize, in essence, the proposed two-phases scenario points to the importance of the localization of the charge in this compound. Indeed, in the Mott insulator, localization by the dopant yields a bad metal, where holes fail to reach the critical screening to completely close the gap, as observed by ARPES. At the same time, disordered regions host the trapped electrons, which show up only in the incoherent transport properties, whereas they are absent from the band structure of the less disordered ones. These electrons appear to unfreeze in the vicinity of the IMT transition. Although an exact modelization certainly appeals for large scale \textit{ab initio} computations yet not available, we believe that our observation of an electron-like contribution to transport could be a test for them.

\section*{Acknowledgements}
We acknowledge support from the Agence Nationale de la Recherche grant ``SOCRATE'' (ANR-15-CE30-0009-01) and LabEx PALM ``ReflectX'' (ANR-10-LABX-0039-PALM).

\bibliography{bibRh} 

\begin{thebibliography}{39}
\expandafter\ifx\csname natexlab\endcsname\relax\def\natexlab#1{#1}\fi
\expandafter\ifx\csname bibnamefont\endcsname\relax
  \def\bibnamefont#1{#1}\fi
\expandafter\ifx\csname bibfnamefont\endcsname\relax
  \def\bibfnamefont#1{#1}\fi
\expandafter\ifx\csname citenamefont\endcsname\relax
  \def\citenamefont#1{#1}\fi
\expandafter\ifx\csname url\endcsname\relax
  \def\url#1{\texttt{#1}}\fi
\expandafter\ifx\csname urlprefix\endcsname\relax\def\urlprefix{URL }\fi
\providecommand{\bibinfo}[2]{#2}
\providecommand{\eprint}[2][]{\url{#2}}

\bibitem[{\citenamefont{Kim et~al.}(2008)\citenamefont{Kim, Jin, Moon, Kim,
  Park, Leem, Yu, Noh, Kim, Oh et~al.}}]{Kim08}
\bibinfo{author}{\bibfnamefont{B.~J.} \bibnamefont{Kim}},
  \bibinfo{author}{\bibfnamefont{H.}~\bibnamefont{Jin}},
  \bibinfo{author}{\bibfnamefont{S.~J.} \bibnamefont{Moon}},
  \bibinfo{author}{\bibfnamefont{J.-Y.} \bibnamefont{Kim}},
  \bibinfo{author}{\bibfnamefont{B.-G.} \bibnamefont{Park}},
  \bibinfo{author}{\bibfnamefont{C.~S.} \bibnamefont{Leem}},
  \bibinfo{author}{\bibfnamefont{J.}~\bibnamefont{Yu}},
  \bibinfo{author}{\bibfnamefont{T.}~\bibnamefont{Noh}},
  \bibinfo{author}{\bibfnamefont{C.}~\bibnamefont{Kim}},
  \bibinfo{author}{\bibfnamefont{S.-J.} \bibnamefont{Oh}},
  \bibnamefont{et~al.}, \bibinfo{journal}{Phys. Rev. Lett.}
  \textbf{\bibinfo{volume}{101}}, \bibinfo{pages}{076402}
  (\bibinfo{year}{2008}).

\bibitem[{\citenamefont{Jackeli and Khaliullin}(2009)}]{Jackeli09}
\bibinfo{author}{\bibfnamefont{G.}~\bibnamefont{Jackeli}} \bibnamefont{and}
  \bibinfo{author}{\bibfnamefont{G.}~\bibnamefont{Khaliullin}},
  \bibinfo{journal}{Phys. Rev. Lett.} \textbf{\bibinfo{volume}{102}},
  \bibinfo{pages}{017205} (\bibinfo{year}{2009}).

\bibitem[{\citenamefont{Korneta et~al.}(2010)\citenamefont{Korneta, Qi,
  Chikara, Parkin, Long, Schlottmann, and Cao}}]{Korneta10}
\bibinfo{author}{\bibfnamefont{O.~B.} \bibnamefont{Korneta}},
  \bibinfo{author}{\bibfnamefont{T.}~\bibnamefont{Qi}},
  \bibinfo{author}{\bibfnamefont{S.}~\bibnamefont{Chikara}},
  \bibinfo{author}{\bibfnamefont{S.}~\bibnamefont{Parkin}},
  \bibinfo{author}{\bibfnamefont{L.~E.~D.} \bibnamefont{Long}},
  \bibinfo{author}{\bibfnamefont{P.}~\bibnamefont{Schlottmann}},
  \bibnamefont{and} \bibinfo{author}{\bibfnamefont{G.}~\bibnamefont{Cao}},
  \bibinfo{journal}{Phys. Rev. B} \textbf{\bibinfo{volume}{82}},
  \bibinfo{pages}{115117} (\bibinfo{year}{2010}).

\bibitem[{\citenamefont{Fruchter et~al.}(2018)\citenamefont{Fruchter, Brouet,
  Colson, Moussy, Forget, and Li}}]{Fruchter18}
\bibinfo{author}{\bibfnamefont{L.}~\bibnamefont{Fruchter}},
  \bibinfo{author}{\bibfnamefont{V.}~\bibnamefont{Brouet}},
  \bibinfo{author}{\bibfnamefont{D.}~\bibnamefont{Colson}},
  \bibinfo{author}{\bibfnamefont{J.-B.} \bibnamefont{Moussy}},
  \bibinfo{author}{\bibfnamefont{A.}~\bibnamefont{Forget}}, \bibnamefont{and}
  \bibinfo{author}{\bibfnamefont{Z.}~\bibnamefont{Li}}, \bibinfo{journal}{J.
  Phys. Chem. Solids} \textbf{\bibinfo{volume}{112}}, \bibinfo{pages}{1}
  (\bibinfo{year}{2018}).

\bibitem[{\citenamefont{Klein and Terasaki}(2008)}]{Klein08}
\bibinfo{author}{\bibfnamefont{Y.}~\bibnamefont{Klein}} \bibnamefont{and}
  \bibinfo{author}{\bibfnamefont{I.}~\bibnamefont{Terasaki}},
  \bibinfo{journal}{J. Phys.: Condens. Matter} \textbf{\bibinfo{volume}{20}},
  \bibinfo{pages}{295201} (\bibinfo{year}{2008}).

\bibitem[{\citenamefont{Ge et~al.}(2011)\citenamefont{Ge, Qi, Korneta, Long,
  Schlottmann, Crummett, and Cao}}]{Ge11}
\bibinfo{author}{\bibfnamefont{M.}~\bibnamefont{Ge}},
  \bibinfo{author}{\bibfnamefont{T.~F.} \bibnamefont{Qi}},
  \bibinfo{author}{\bibfnamefont{B.}~\bibnamefont{Korneta}},
  \bibinfo{author}{\bibfnamefont{E.~D.} \bibnamefont{Long}},
  \bibinfo{author}{\bibfnamefont{P.}~\bibnamefont{Schlottmann}},
  \bibinfo{author}{\bibfnamefont{W.~P.} \bibnamefont{Crummett}},
  \bibnamefont{and} \bibinfo{author}{\bibfnamefont{G.}~\bibnamefont{Cao}},
  \bibinfo{journal}{Phys. Rev. B} \textbf{\bibinfo{volume}{84}},
  \bibinfo{pages}{100402(R)} (\bibinfo{year}{2011}).

\bibitem[{\citenamefont{Brouet et~al.}(2015)\citenamefont{Brouet, Mansart,
  Perfetti, Piovera, Vobornik, F\`evre, Bertran, Riggs, Shapiro, Giraldo-Gallo
  et~al.}}]{Brouet15}
\bibinfo{author}{\bibfnamefont{V.}~\bibnamefont{Brouet}},
  \bibinfo{author}{\bibfnamefont{J.}~\bibnamefont{Mansart}},
  \bibinfo{author}{\bibfnamefont{L.}~\bibnamefont{Perfetti}},
  \bibinfo{author}{\bibfnamefont{C.}~\bibnamefont{Piovera}},
  \bibinfo{author}{\bibfnamefont{I.}~\bibnamefont{Vobornik}},
  \bibinfo{author}{\bibfnamefont{P.~L.} \bibnamefont{F\`evre}},
  \bibinfo{author}{\bibfnamefont{F.}~\bibnamefont{Bertran}},
  \bibinfo{author}{\bibfnamefont{S.}~\bibnamefont{Riggs}},
  \bibinfo{author}{\bibfnamefont{M.~C.} \bibnamefont{Shapiro}},
  \bibinfo{author}{\bibfnamefont{P.}~\bibnamefont{Giraldo-Gallo}},
  \bibnamefont{et~al.}, \bibinfo{journal}{Phys. Rev. B}
  \textbf{\bibinfo{volume}{92}}, \bibinfo{pages}{081117(R)}
  (\bibinfo{year}{2015}).

\bibitem[{\citenamefont{Yan et~al.}(2015)\citenamefont{Yan, Ren, Xu, Xie, Tao,
  Choi, Lee, Choi, Zhang, and Feng}}]{Yan15}
\bibinfo{author}{\bibfnamefont{Y.~J.} \bibnamefont{Yan}},
  \bibinfo{author}{\bibfnamefont{M.~Q.} \bibnamefont{Ren}},
  \bibinfo{author}{\bibfnamefont{H.~C.} \bibnamefont{Xu}},
  \bibinfo{author}{\bibfnamefont{B.~P.} \bibnamefont{Xie}},
  \bibinfo{author}{\bibfnamefont{R.}~\bibnamefont{Tao}},
  \bibinfo{author}{\bibfnamefont{H.~Y.} \bibnamefont{Choi}},
  \bibinfo{author}{\bibfnamefont{N.}~\bibnamefont{Lee}},
  \bibinfo{author}{\bibfnamefont{Y.~J.} \bibnamefont{Choi}},
  \bibinfo{author}{\bibfnamefont{T.}~\bibnamefont{Zhang}}, \bibnamefont{and}
  \bibinfo{author}{\bibfnamefont{D.~L.} \bibnamefont{Feng}},
  \bibinfo{journal}{Phys. Rev. X} \textbf{\bibinfo{volume}{5}},
  \bibinfo{pages}{041018} (\bibinfo{year}{2015}).

\bibitem[{\citenamefont{Baumberger et~al.}(2006)\citenamefont{Baumberger,
  Ingle, Meevasana, Shen, Lu, Perry, Mackenzie, Hussain, Singh, and
  Shen}}]{Baumberger06}
\bibinfo{author}{\bibfnamefont{F.}~\bibnamefont{Baumberger}},
  \bibinfo{author}{\bibfnamefont{N.}~\bibnamefont{Ingle}},
  \bibinfo{author}{\bibfnamefont{W.}~\bibnamefont{Meevasana}},
  \bibinfo{author}{\bibfnamefont{K.}~\bibnamefont{Shen}},
  \bibinfo{author}{\bibfnamefont{D.}~\bibnamefont{Lu}},
  \bibinfo{author}{\bibfnamefont{R.}~\bibnamefont{Perry}},
  \bibinfo{author}{\bibfnamefont{A.}~\bibnamefont{Mackenzie}},
  \bibinfo{author}{\bibfnamefont{Z.}~\bibnamefont{Hussain}},
  \bibinfo{author}{\bibfnamefont{D.}~\bibnamefont{Singh}}, \bibnamefont{and}
  \bibinfo{author}{\bibfnamefont{Z.-X.} \bibnamefont{Shen}},
  \bibinfo{journal}{Phys. Rev. Lett.} \textbf{\bibinfo{volume}{96}},
  \bibinfo{pages}{246402} (\bibinfo{year}{2006}).

\bibitem[{\citenamefont{Martins et~al.}(2011)\citenamefont{Martins, Aichhorn,
  Vaugier, and Biermann}}]{Martins11}
\bibinfo{author}{\bibfnamefont{C.}~\bibnamefont{Martins}},
  \bibinfo{author}{\bibfnamefont{M.}~\bibnamefont{Aichhorn}},
  \bibinfo{author}{\bibfnamefont{L.}~\bibnamefont{Vaugier}}, \bibnamefont{and}
  \bibinfo{author}{\bibfnamefont{S.}~\bibnamefont{Biermann}},
  \bibinfo{journal}{Phys. Rev. Lett.} \textbf{\bibinfo{volume}{107}},
  \bibinfo{pages}{266404} (\bibinfo{year}{2011}).

\bibitem[{\citenamefont{Qi et~al.}(2012)\citenamefont{Qi, Korneta, L.~Li, Cao,
  Wan, Schlottmann, Kaul, , and Cao}}]{Qi12}
\bibinfo{author}{\bibfnamefont{T.~F.} \bibnamefont{Qi}},
  \bibinfo{author}{\bibfnamefont{O.~B.} \bibnamefont{Korneta}},
  \bibinfo{author}{\bibfnamefont{K.~B.} \bibnamefont{L.~Li}},
  \bibinfo{author}{\bibfnamefont{V.~S.} \bibnamefont{Cao}},
  \bibinfo{author}{\bibfnamefont{X.}~\bibnamefont{Wan}},
  \bibinfo{author}{\bibfnamefont{P.}~\bibnamefont{Schlottmann}},
  \bibinfo{author}{\bibfnamefont{R.~K.} \bibnamefont{Kaul}}, ,
  \bibnamefont{and} \bibinfo{author}{\bibfnamefont{G.}~\bibnamefont{Cao}},
  \bibinfo{journal}{Phys. Rev. B} \textbf{\bibinfo{volume}{86}},
  \bibinfo{pages}{125105} (\bibinfo{year}{2012}).

\bibitem[{\citenamefont{Clancy et~al.}(2014)\citenamefont{Clancy, Lupascu,
  Gretarsson, Islam, Hu, Casa, Nelson, LaMarra, Cao, and Kim}}]{Clancy14}
\bibinfo{author}{\bibfnamefont{J.~P.} \bibnamefont{Clancy}},
  \bibinfo{author}{\bibfnamefont{A.}~\bibnamefont{Lupascu}},
  \bibinfo{author}{\bibfnamefont{H.}~\bibnamefont{Gretarsson}},
  \bibinfo{author}{\bibfnamefont{Z.}~\bibnamefont{Islam}},
  \bibinfo{author}{\bibfnamefont{Y.~F.} \bibnamefont{Hu}},
  \bibinfo{author}{\bibfnamefont{D.}~\bibnamefont{Casa}},
  \bibinfo{author}{\bibfnamefont{C.~S.} \bibnamefont{Nelson}},
  \bibinfo{author}{\bibfnamefont{S.~C.} \bibnamefont{LaMarra}},
  \bibinfo{author}{\bibfnamefont{G.}~\bibnamefont{Cao}}, \bibnamefont{and}
  \bibinfo{author}{\bibfnamefont{Y.-J.} \bibnamefont{Kim}},
  \bibinfo{journal}{Phys. Rev. B} \textbf{\bibinfo{volume}{89}},
  \bibinfo{pages}{054409} (\bibinfo{year}{2014}).

\bibitem[{\citenamefont{Klein and Terasaki}(2009)}]{Klein09}
\bibinfo{author}{\bibfnamefont{Y.}~\bibnamefont{Klein}} \bibnamefont{and}
  \bibinfo{author}{\bibfnamefont{I.}~\bibnamefont{Terasaki}},
  \bibinfo{journal}{Journal of electronic materials}
  \textbf{\bibinfo{volume}{38}}, \bibinfo{pages}{1331} (\bibinfo{year}{2009}).

\bibitem[{\citenamefont{Lee et~al.}(2012)\citenamefont{Lee, Krockenberger,
  Takahashi, Kawasaki, and Tokura}}]{Lee12}
\bibinfo{author}{\bibfnamefont{J.~S.} \bibnamefont{Lee}},
  \bibinfo{author}{\bibfnamefont{Y.}~\bibnamefont{Krockenberger}},
  \bibinfo{author}{\bibfnamefont{K.~S.} \bibnamefont{Takahashi}},
  \bibinfo{author}{\bibfnamefont{M.}~\bibnamefont{Kawasaki}}, \bibnamefont{and}
  \bibinfo{author}{\bibfnamefont{Y.}~\bibnamefont{Tokura}},
  \bibinfo{journal}{Phys. Rev. B} \textbf{\bibinfo{volume}{85}},
  \bibinfo{pages}{035101} (\bibinfo{year}{2012}).

\bibitem[{\citenamefont{Sohn et~al.}(2014)\citenamefont{Sohn, Cho, Kuo,
  Sandilands, Qi, Cao, and Noh}}]{Sohn14}
\bibinfo{author}{\bibfnamefont{C.~H.} \bibnamefont{Sohn}},
  \bibinfo{author}{\bibfnamefont{D.-Y.} \bibnamefont{Cho}},
  \bibinfo{author}{\bibfnamefont{C.-T.} \bibnamefont{Kuo}},
  \bibinfo{author}{\bibfnamefont{L.~J.} \bibnamefont{Sandilands}},
  \bibinfo{author}{\bibfnamefont{T.~F.} \bibnamefont{Qi}},
  \bibinfo{author}{\bibfnamefont{G.}~\bibnamefont{Cao}}, \bibnamefont{and}
  \bibinfo{author}{\bibfnamefont{T.~W.} \bibnamefont{Noh}},
  \bibinfo{journal}{Sci. Rep.} \textbf{\bibinfo{volume}{6}},
  \bibinfo{pages}{23856} (\bibinfo{year}{2014}).

\bibitem[{\citenamefont{Cao et~al.}(2016)\citenamefont{Cao, Wang, Waugh, Reber,
  Li, Zhou, Parham, Plumb, Rotenberg, Bostwick et~al.}}]{Cao16}
\bibinfo{author}{\bibfnamefont{Y.}~\bibnamefont{Cao}},
  \bibinfo{author}{\bibfnamefont{Q.}~\bibnamefont{Wang}},
  \bibinfo{author}{\bibfnamefont{J.~A.} \bibnamefont{Waugh}},
  \bibinfo{author}{\bibfnamefont{T.~J.} \bibnamefont{Reber}},
  \bibinfo{author}{\bibfnamefont{H.}~\bibnamefont{Li}},
  \bibinfo{author}{\bibfnamefont{X.}~\bibnamefont{Zhou}},
  \bibinfo{author}{\bibfnamefont{S.}~\bibnamefont{Parham}},
  \bibinfo{author}{\bibfnamefont{N.~C.} \bibnamefont{Plumb}},
  \bibinfo{author}{\bibfnamefont{E.}~\bibnamefont{Rotenberg}},
  \bibinfo{author}{\bibfnamefont{A.}~\bibnamefont{Bostwick}},
  \bibnamefont{et~al.}, \bibinfo{journal}{Nature Com.}
  \textbf{\bibinfo{volume}{7}}, \bibinfo{pages}{11367} (\bibinfo{year}{2016}).

\bibitem[{\citenamefont{Liu et~al.}(2016)\citenamefont{Liu, Reticcioli, Kim,
  Continenza, Kresse, Sarma, Chen, and Franchini}}]{Liu16}
\bibinfo{author}{\bibfnamefont{P.}~\bibnamefont{Liu}},
  \bibinfo{author}{\bibfnamefont{M.}~\bibnamefont{Reticcioli}},
  \bibinfo{author}{\bibfnamefont{B.}~\bibnamefont{Kim}},
  \bibinfo{author}{\bibfnamefont{A.}~\bibnamefont{Continenza}},
  \bibinfo{author}{\bibfnamefont{G.}~\bibnamefont{Kresse}},
  \bibinfo{author}{\bibfnamefont{D.~D.} \bibnamefont{Sarma}},
  \bibinfo{author}{\bibfnamefont{X.-Q.} \bibnamefont{Chen}}, \bibnamefont{and}
  \bibinfo{author}{\bibfnamefont{C.}~\bibnamefont{Franchini}},
  \bibinfo{journal}{Phys. Rev. B} \textbf{\bibinfo{volume}{94}},
  \bibinfo{pages}{195145} (\bibinfo{year}{2016}).

\bibitem[{\citenamefont{Xu et~al.}(2020)\citenamefont{Xu, Marsik, Sheveleva,
  Lyzwa, Louat, Brouet, Munzar, and Bernhard}}]{Xu20}
\bibinfo{author}{\bibfnamefont{B.}~\bibnamefont{Xu}},
  \bibinfo{author}{\bibfnamefont{P.}~\bibnamefont{Marsik}},
  \bibinfo{author}{\bibfnamefont{E.}~\bibnamefont{Sheveleva}},
  \bibinfo{author}{\bibfnamefont{F.}~\bibnamefont{Lyzwa}},
  \bibinfo{author}{\bibfnamefont{A.}~\bibnamefont{Louat}},
  \bibinfo{author}{\bibfnamefont{V.}~\bibnamefont{Brouet}},
  \bibinfo{author}{\bibfnamefont{D.}~\bibnamefont{Munzar}}, \bibnamefont{and}
  \bibinfo{author}{\bibfnamefont{C.}~\bibnamefont{Bernhard}},
  \bibinfo{journal}{Phys. Rev. Lett.} \textbf{\bibinfo{volume}{124}},
  \bibinfo{pages}{027402} (\bibinfo{year}{2020}).

\bibitem[{\citenamefont{Chikara et~al.}(2017)\citenamefont{Chikara, Fabbris,
  Terzic, Cao, Khomskii, and Haskel}}]{Chikara17}
\bibinfo{author}{\bibfnamefont{S.}~\bibnamefont{Chikara}},
  \bibinfo{author}{\bibfnamefont{G.}~\bibnamefont{Fabbris}},
  \bibinfo{author}{\bibfnamefont{J.}~\bibnamefont{Terzic}},
  \bibinfo{author}{\bibfnamefont{G.}~\bibnamefont{Cao}},
  \bibinfo{author}{\bibfnamefont{D.}~\bibnamefont{Khomskii}}, \bibnamefont{and}
  \bibinfo{author}{\bibfnamefont{D.}~\bibnamefont{Haskel}},
  \bibinfo{journal}{Phys. Rev. B} \textbf{\bibinfo{volume}{95}},
  \bibinfo{pages}{060407(R)} (\bibinfo{year}{2017}).

\bibitem[{\citenamefont{Louat et~al.}(2018)\citenamefont{Louat, Bert,
  Serrier-Garcia, Bertran, F\`{e}vre, Rault, and Brouet}}]{Louat18}
\bibinfo{author}{\bibfnamefont{A.}~\bibnamefont{Louat}},
  \bibinfo{author}{\bibfnamefont{F.}~\bibnamefont{Bert}},
  \bibinfo{author}{\bibfnamefont{L.}~\bibnamefont{Serrier-Garcia}},
  \bibinfo{author}{\bibfnamefont{F.}~\bibnamefont{Bertran}},
  \bibinfo{author}{\bibfnamefont{P.~L.} \bibnamefont{F\`{e}vre}},
  \bibinfo{author}{\bibfnamefont{J.~E.} \bibnamefont{Rault}}, \bibnamefont{and}
  \bibinfo{author}{\bibfnamefont{V.}~\bibnamefont{Brouet}},
  \bibinfo{journal}{Phys. Rev. B} \textbf{\bibinfo{volume}{97}},
  \bibinfo{pages}{161109(R)} (\bibinfo{year}{2018}).

\bibitem[{\citenamefont{Louat et~al.}(2019)\citenamefont{Louat, Lenz, Biermann,
  Martins, Bertran, F\`{e}vre, Rault, Bert, and Brouet}}]{Louat19}
\bibinfo{author}{\bibfnamefont{A.}~\bibnamefont{Louat}},
  \bibinfo{author}{\bibfnamefont{B.}~\bibnamefont{Lenz}},
  \bibinfo{author}{\bibfnamefont{S.}~\bibnamefont{Biermann}},
  \bibinfo{author}{\bibfnamefont{C.}~\bibnamefont{Martins}},
  \bibinfo{author}{\bibfnamefont{F.}~\bibnamefont{Bertran}},
  \bibinfo{author}{\bibfnamefont{P.~L.} \bibnamefont{F\`{e}vre}},
  \bibinfo{author}{\bibfnamefont{J.~E.} \bibnamefont{Rault}},
  \bibinfo{author}{\bibfnamefont{F.}~\bibnamefont{Bert}}, \bibnamefont{and}
  \bibinfo{author}{\bibfnamefont{V.}~\bibnamefont{Brouet}},
  \bibinfo{journal}{Phys. Rev. B} \textbf{\bibinfo{volume}{100}},
  \bibinfo{pages}{205135} (\bibinfo{year}{2019}).

\bibitem[{\citenamefont{Kim et~al.}(2009)\citenamefont{Kim, Ohsumi, Komesu,
  Sakai, Morita, Takagi, and Arima}}]{Kim09}
\bibinfo{author}{\bibfnamefont{B.~J.} \bibnamefont{Kim}},
  \bibinfo{author}{\bibfnamefont{H.}~\bibnamefont{Ohsumi}},
  \bibinfo{author}{\bibfnamefont{T.}~\bibnamefont{Komesu}},
  \bibinfo{author}{\bibfnamefont{S.}~\bibnamefont{Sakai}},
  \bibinfo{author}{\bibfnamefont{T.}~\bibnamefont{Morita}},
  \bibinfo{author}{\bibfnamefont{H.}~\bibnamefont{Takagi}}, \bibnamefont{and}
  \bibinfo{author}{\bibfnamefont{T.}~\bibnamefont{Arima}},
  \bibinfo{journal}{Science} \textbf{\bibinfo{volume}{323}},
  \bibinfo{pages}{1329} (\bibinfo{year}{2009}).

\bibitem[{\citenamefont{Herring}(1954)}]{Herring54}
\bibinfo{author}{\bibfnamefont{C.}~\bibnamefont{Herring}},
  \bibinfo{journal}{Phys. Rev.} \textbf{\bibinfo{volume}{96}},
  \bibinfo{pages}{1163} (\bibinfo{year}{1954}).

\bibitem[{\citenamefont{Ishii et~al.}(1987)\citenamefont{Ishii, Sato, Kanazawa,
  Takagi, Uchida, Kitazawa, Kishio, Fueki, and Tanaka}}]{Ishii87}
\bibinfo{author}{\bibfnamefont{H.}~\bibnamefont{Ishii}},
  \bibinfo{author}{\bibfnamefont{H.}~\bibnamefont{Sato}},
  \bibinfo{author}{\bibfnamefont{N.}~\bibnamefont{Kanazawa}},
  \bibinfo{author}{\bibfnamefont{H.}~\bibnamefont{Takagi}},
  \bibinfo{author}{\bibfnamefont{S.}~\bibnamefont{Uchida}},
  \bibinfo{author}{\bibfnamefont{K.}~\bibnamefont{Kitazawa}},
  \bibinfo{author}{\bibfnamefont{K.}~\bibnamefont{Kishio}},
  \bibinfo{author}{\bibfnamefont{K.}~\bibnamefont{Fueki}}, \bibnamefont{and}
  \bibinfo{author}{\bibfnamefont{S.}~\bibnamefont{Tanaka}},
  \bibinfo{journal}{Physica} \textbf{\bibinfo{volume}{148B}},
  \bibinfo{pages}{419} (\bibinfo{year}{1987}).

\bibitem[{\citenamefont{Dugdale}(1995)}]{Dugdale95}
\bibinfo{author}{\bibfnamefont{J.~S.} \bibnamefont{Dugdale}},
  \emph{\bibinfo{title}{The Electrical Properties of Disordered Metals}}
  (\bibinfo{publisher}{Cambridge University Press}, \bibinfo{year}{1995}).

\bibitem[{\citenamefont{Badoux et~al.}(2016)\citenamefont{Badoux, Tabis,
  Lalibert\'{e}, Grissonnanche, Vignolle, Vignolles, B\'{e}ard, Bonn, Hardy,
  Liang et~al.}}]{Badoux16}
\bibinfo{author}{\bibfnamefont{S.}~\bibnamefont{Badoux}},
  \bibinfo{author}{\bibfnamefont{W.}~\bibnamefont{Tabis}},
  \bibinfo{author}{\bibfnamefont{F.}~\bibnamefont{Lalibert\'{e}}},
  \bibinfo{author}{\bibfnamefont{G.}~\bibnamefont{Grissonnanche}},
  \bibinfo{author}{\bibfnamefont{B.}~\bibnamefont{Vignolle}},
  \bibinfo{author}{\bibfnamefont{D.}~\bibnamefont{Vignolles}},
  \bibinfo{author}{\bibfnamefont{J.}~\bibnamefont{B\'{e}ard}},
  \bibinfo{author}{\bibfnamefont{D.~A.} \bibnamefont{Bonn}},
  \bibinfo{author}{\bibfnamefont{W.~N.} \bibnamefont{Hardy}},
  \bibinfo{author}{\bibfnamefont{R.}~\bibnamefont{Liang}},
  \bibnamefont{et~al.}, \bibinfo{journal}{Nature}
  \textbf{\bibinfo{volume}{531}}, \bibinfo{pages}{210} (\bibinfo{year}{2016}).

\bibitem[{\citenamefont{Nelson et~al.}(2020)\citenamefont{Nelson, Parzyck,
  Faeth1, Kawasaki, Schlom, and Shen}}]{Nelson20}
\bibinfo{author}{\bibfnamefont{J.~N.} \bibnamefont{Nelson}},
  \bibinfo{author}{\bibfnamefont{C.~T.} \bibnamefont{Parzyck}},
  \bibinfo{author}{\bibfnamefont{B.~D.} \bibnamefont{Faeth1}},
  \bibinfo{author}{\bibfnamefont{J.~K.} \bibnamefont{Kawasaki}},
  \bibinfo{author}{\bibfnamefont{D.~G.} \bibnamefont{Schlom}},
  \bibnamefont{and} \bibinfo{author}{\bibfnamefont{K.~M.} \bibnamefont{Shen}},
  \bibinfo{journal}{Nature Com.} \textbf{\bibinfo{volume}{11}},
  \bibinfo{pages}{2597} (\bibinfo{year}{2020}).

\bibitem[{\citenamefont{Ashcroft and Mermin}(1976)}]{Ashcroft}
\bibinfo{author}{\bibfnamefont{N.~W.} \bibnamefont{Ashcroft}} \bibnamefont{and}
  \bibinfo{author}{\bibfnamefont{N.~D.} \bibnamefont{Mermin}},
  \emph{\bibinfo{title}{Solid State Physics}} (\bibinfo{publisher}{Saunders
  College}, \bibinfo{year}{1976}).

\bibitem[{\citenamefont{Chikara et~al.}(2015)\citenamefont{Chikara, Haskel,
  Sim, Kim, Chen, Fabbris, Veiga, Souza-Neto, Terzic, Butrouna
  et~al.}}]{Chikara15}
\bibinfo{author}{\bibfnamefont{S.}~\bibnamefont{Chikara}},
  \bibinfo{author}{\bibfnamefont{D.}~\bibnamefont{Haskel}},
  \bibinfo{author}{\bibfnamefont{J.-H.} \bibnamefont{Sim}},
  \bibinfo{author}{\bibfnamefont{H.-S.} \bibnamefont{Kim}},
  \bibinfo{author}{\bibfnamefont{C.-C.} \bibnamefont{Chen}},
  \bibinfo{author}{\bibfnamefont{G.}~\bibnamefont{Fabbris}},
  \bibinfo{author}{\bibfnamefont{L.~S.~I.} \bibnamefont{Veiga}},
  \bibinfo{author}{\bibfnamefont{N.~M.} \bibnamefont{Souza-Neto}},
  \bibinfo{author}{\bibfnamefont{J.}~\bibnamefont{Terzic}},
  \bibinfo{author}{\bibfnamefont{K.}~\bibnamefont{Butrouna}},
  \bibnamefont{et~al.}, \bibinfo{journal}{Phys. Rev. B}
  \textbf{\bibinfo{volume}{92}}, \bibinfo{pages}{081114(R)}
  (\bibinfo{year}{2015}).

\bibitem[{\citenamefont{Levy and Bergman}(1992)}]{Levy92}
\bibinfo{author}{\bibfnamefont{O.}~\bibnamefont{Levy}} \bibnamefont{and}
  \bibinfo{author}{\bibfnamefont{D.~J.} \bibnamefont{Bergman}},
  \bibinfo{journal}{J. Phys. A: Math. Gen.} \textbf{\bibinfo{volume}{25}},
  \bibinfo{pages}{1875} (\bibinfo{year}{1992}).

\bibitem[{\citenamefont{Kleber}(2005)}]{Kleber05}
\bibinfo{author}{\bibfnamefont{X.}~\bibnamefont{Kleber}},
  \bibinfo{journal}{Modelling Simul. Mater. Sci. Eng.}
  \textbf{\bibinfo{volume}{14}}, \bibinfo{pages}{21} (\bibinfo{year}{2005}).

\bibitem[{\citenamefont{Johnston}(2002)}]{Johnston02}
\bibinfo{author}{\bibfnamefont{R.}~\bibnamefont{Johnston}},
  \emph{\bibinfo{title}{Atomic and Molecular Clusters}}
  (\bibinfo{publisher}{Taylor \& Francis}, \bibinfo{year}{2002}).

\bibitem[{\citenamefont{Stauffer and Aharony}(1991)}]{Stauffer91}
\bibinfo{author}{\bibfnamefont{D.}~\bibnamefont{Stauffer}} \bibnamefont{and}
  \bibinfo{author}{\bibfnamefont{A.}~\bibnamefont{Aharony}},
  \emph{\bibinfo{title}{Introduction to Percolation Theory}}
  (\bibinfo{publisher}{Taylor \& Francis}, \bibinfo{year}{1991}).

\bibitem[{\citenamefont{Liu et~al.}(2008)\citenamefont{Liu, Antonov, Jepsen,
  and Andersen}}]{Liu08}
\bibinfo{author}{\bibfnamefont{G.-Q.} \bibnamefont{Liu}},
  \bibinfo{author}{\bibfnamefont{V.~N.} \bibnamefont{Antonov}},
  \bibinfo{author}{\bibfnamefont{O.}~\bibnamefont{Jepsen}}, \bibnamefont{and}
  \bibinfo{author}{\bibfnamefont{O.~K.} \bibnamefont{Andersen}},
  \bibinfo{journal}{Phys. Rev. Lett.} \textbf{\bibinfo{volume}{101}},
  \bibinfo{pages}{026408} (\bibinfo{year}{2008}).

\bibitem[{\citenamefont{Blaha et~al.}(1990)\citenamefont{Blaha, Schwarz,
  Sorantin, and Trickey}}]{Blaha90}
\bibinfo{author}{\bibfnamefont{P.}~\bibnamefont{Blaha}},
  \bibinfo{author}{\bibfnamefont{K.}~\bibnamefont{Schwarz}},
  \bibinfo{author}{\bibfnamefont{P.}~\bibnamefont{Sorantin}}, \bibnamefont{and}
  \bibinfo{author}{\bibfnamefont{S.}~\bibnamefont{Trickey}},
  \bibinfo{journal}{Comput. Phys. Commun.} \textbf{\bibinfo{volume}{59}},
  \bibinfo{pages}{399} (\bibinfo{year}{1990}).

\bibitem[{\citenamefont{Ahn et~al.}(2014)\citenamefont{Ahn, Lee, and
  Kunes}}]{Ahn15}
\bibinfo{author}{\bibfnamefont{K.-H.} \bibnamefont{Ahn}},
  \bibinfo{author}{\bibfnamefont{K.-W.} \bibnamefont{Lee}}, \bibnamefont{and}
  \bibinfo{author}{\bibfnamefont{J.}~\bibnamefont{Kunes}}, \bibinfo{journal}{J.
  Phys.: Condens. Matter} \textbf{\bibinfo{volume}{27}},
  \bibinfo{pages}{085602} (\bibinfo{year}{2014}).

\bibitem[{\citenamefont{Sasioglu et~al.}(2011)\citenamefont{Sasioglu,
  Friedrich, and Blugel}}]{Sasioglu11}
\bibinfo{author}{\bibfnamefont{E.}~\bibnamefont{Sasioglu}},
  \bibinfo{author}{\bibfnamefont{C.}~\bibnamefont{Friedrich}},
  \bibnamefont{and} \bibinfo{author}{\bibfnamefont{S.}~\bibnamefont{Blugel}},
  \bibinfo{journal}{Phys. Rev. B} \textbf{\bibinfo{volume}{83}},
  \bibinfo{pages}{121101(R)} (\bibinfo{year}{2011}).

\bibitem[{\citenamefont{Shimura et~al.}(1994)\citenamefont{Shimura, Itoh,
  Inaguma, and Naksmura}}]{Shimura94}
\bibinfo{author}{\bibfnamefont{T.}~\bibnamefont{Shimura}},
  \bibinfo{author}{\bibfnamefont{M.}~\bibnamefont{Itoh}},
  \bibinfo{author}{\bibfnamefont{Y.}~\bibnamefont{Inaguma}}, \bibnamefont{and}
  \bibinfo{author}{\bibfnamefont{T.}~\bibnamefont{Naksmura}},
  \bibinfo{journal}{Phys. Rev. B} \textbf{\bibinfo{volume}{49}},
  \bibinfo{pages}{5591} (\bibinfo{year}{1994}).

\bibitem[{\citenamefont{Kwon et~al.}(2019)\citenamefont{Kwon, Kim, Song,
  Yoshida, Denlinger, Kyung, and Kim}}]{Kwon19}
\bibinfo{author}{\bibfnamefont{J.}~\bibnamefont{Kwon}},
  \bibinfo{author}{\bibfnamefont{M.}~\bibnamefont{Kim}},
  \bibinfo{author}{\bibfnamefont{D.}~\bibnamefont{Song}},
  \bibinfo{author}{\bibfnamefont{Y.}~\bibnamefont{Yoshida}},
  \bibinfo{author}{\bibfnamefont{J.}~\bibnamefont{Denlinger}},
  \bibinfo{author}{\bibfnamefont{W.}~\bibnamefont{Kyung}}, \bibnamefont{and}
  \bibinfo{author}{\bibfnamefont{C.}~\bibnamefont{Kim}},
  \bibinfo{journal}{Phys. Rev. Lett.} \textbf{\bibinfo{volume}{123}},
  \bibinfo{pages}{106401} (\bibinfo{year}{2019}).

\end{thebibliography}

\end{document}